\documentclass[twocolumn,prl,showpacs]{revtex4}

\usepackage{graphicx}
\usepackage{dcolumn}
\usepackage{bm}

\begin{document}

\title{Universal Features of the Time Evolution of Evanescent Modes 
 in a Left-Handed Perfect Lens.}

\author{ G. G\'omez-Santos }

\affiliation{Departamento~de~F\'{\i}sica~de~la~Materia~Condensada 
 and Instituto~Nicol\'as~Cabrera,~Universidad~Aut\'onoma~de~Madrid,~28049~Madrid,~Spain.}

\begin{abstract}
The time evolution of evanescent modes in Pendry's perfect lens proposal for 
ideally lossless and homogeneous, left-handed materials is analyzed. We show
that time development of sub-wavelength resolution exhibits universal 
features, independent of model details. This is due to the unavoidable 
near-degeneracy of surface electromagnetic modes in the deep sub-wavelength 
region. By means of a mechanical analog, it is shown that an intrinsic time
scale (missed in stationary studies)  has to be associated with any desired 
lateral resolution. A time-dependent cut-off length emerges, removing  the 
problem of divergences  claimed to invalidate Pendry's proposal.
\end{abstract}

\maketitle


  Long ago, Veselago \cite{veselago68} pointed out that very unusual properties, such as negative
  refraction,
  would be exhibited by materials with negative refraction index.
  Recently, Pendry \cite{pendry00} has claimed that those so called left-handed (LH)  materials with 
  \hbox{$ \epsilon(\omega_0)=\mu(\omega_0)=-1$}, can act as perfect
  lenses with, ideally, arbitrary sub-wavelength resolution. In addition to its genuine
  conceptual importance, this proposal has attracted much attention due to the practical
  realization of man-made
  materials expected to be left-handed \cite{padilla00}, 
  where negative refraction has been claimed to be observed \cite{shelby01}.
  The challenge to conventional ideas conveyed in Pendry's work has fueled a 
  heated debate that has contributed to sharpen the issue, if not to settle it 
  \cite{hooft01,valanju02,garcia02,pendry02}.
  
    The criticism expressed 
   by Garc\'{\i}a and N-Vesperinas \cite{garcia02} (see also  Refs. \cite{hooft01,
   pokrovsky02}) seems of particular importance,
    for it would imply  that some
  sort of fundamental violation of physical laws is unavoidable in Pendry's perfect
  lens. The idea is that the necessary amplification of evanescent modes
  would turn a square-integrable incoming
  wave into a non normalizable signal, something deemed unacceptable by the authors of
  Refs. \cite{garcia02,hooft01,pokrovsky02}. Pendry \cite{pendry02} has replied that losses, 
  unavoidable in the real world, would
  provide a natural cut-off, forbidding the amplification of large wavevectors
  (though  degrading the perfect resolution). Haldane \cite{haldane02} 
  has put the blame of divergences  on the 
  role played by surface  modes (polaritons) \cite{ruppin00} localized at the interfaces of the LH
  material: amplification of evanescent modes is a gift of those modes \cite{pendry00}. For a 
  homogeneous material, those surface modes exist and become dispersionless for
  increasing wavevectors, being the unphysical absence of an intrinsic  cut-off
  length which
  causes the pathologies \cite{haldane02}. Any  realization of a LH material (think of the
  composite nature of proposed LH materials \cite{shelby01} or  photonic crystals \cite{luo02})
  must  
  have a natural length  below which the homogeneous description fails, providing
  the necessary  cure to divergences at the price of lower  resolution.
  
   Although Pendry's and Haldane's escapes from divergences are certainly safe, they seem to suggest that the
  textbook idealization of lossless and homogeneous media, that works so well otherwise
  \cite{jackson}, is
  fundamentally flawed when applied to a material that happens to satisfy
  $\epsilon(\omega_0)=\mu(\omega_0)=-1$, 
  at some frequency $\omega_0$.  Apparently, only after the inclusion on the real world constraints 
  (losses and/or small-distance structure) could the proposal be made 
  physically acceptable. I find this state of affairs very unsatisfactory, and it is the purpose of this
  paper to show that, {\em even within the self-imposed idealizations of a lossless 
  (for $\omega=\omega_0$) and purely
  homogeneous, left-handed material, Pendry's perfect lens proposal is correct}.

  To this end, we will consider the time evolution \cite{ziolkowski01,anantha02} of 
  sub-wavelength features.
  Time development requires the study of the
   frequency dispersion of a particular model (or material), apparently preventing us from
  drawing general conclusions. However, it will be shown that the structure of 
  the relevant magnitudes in the
  immediate vicinity of the target frequency is dominated by the surface polariton modes
  \cite{ruppin00,haldane02}. These modes 
  show  
  universal features for any LH material at long wavevectors, therefore allowing us
  to extract general conclusions. A key point will be the identification of a time scale 
  for any intended length resolution. This time scale, 
  linked to the near-degeneracy of surface modes and missed in stationary studies, 
  will provide the
  clue for overcoming the problem of divergences.

\begin{figure} 
\includegraphics [clip,width=8.cm]{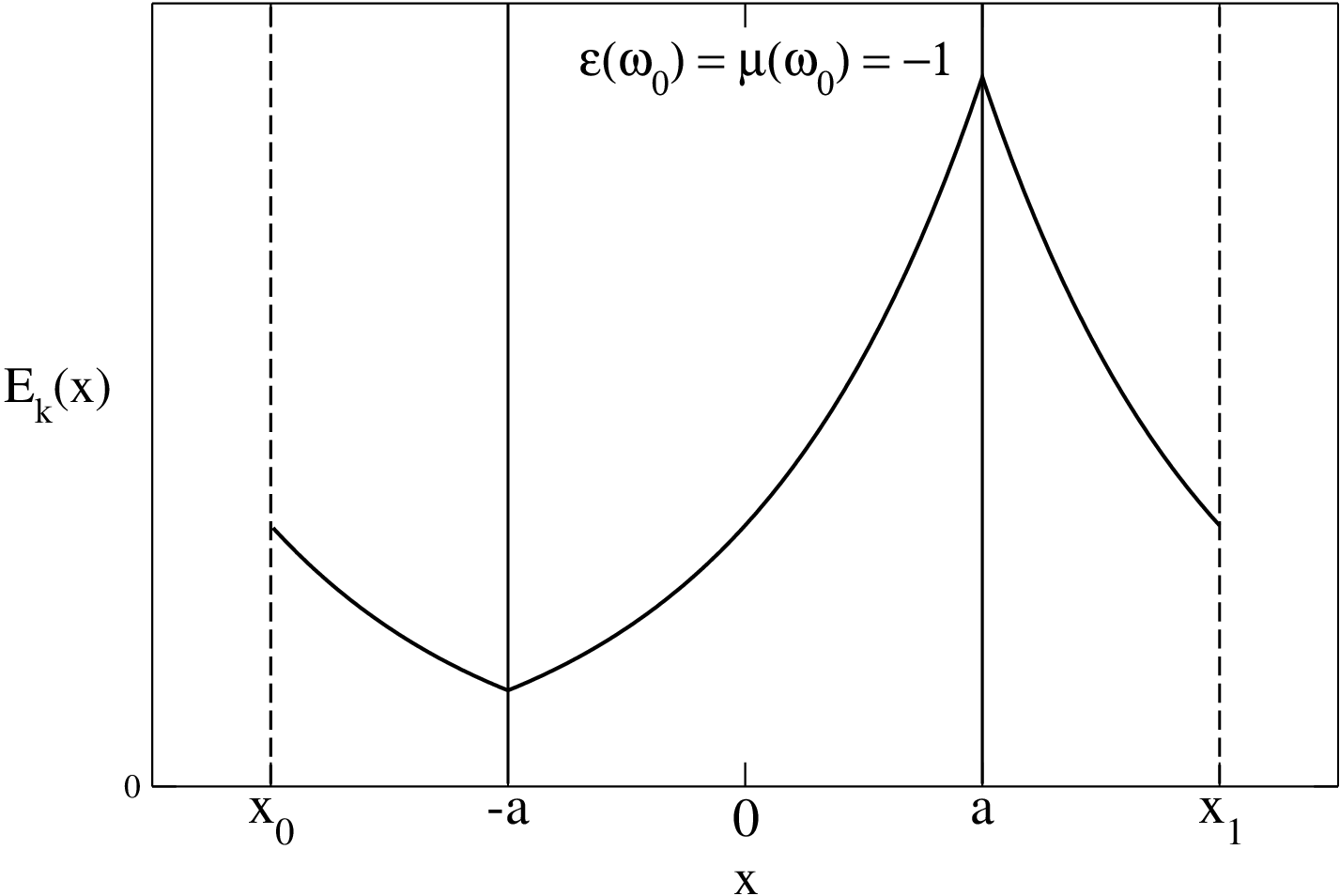}
\caption{Geometry of the problem: slab of LH material between planes $x =\pm a$,
with source plane at $x_0$, and image plane at $x_1$. The curve of 
$ E_{k}(x)$ illustrates  perfect restoration of evanescent fields when
$x_0=-2a$ and $x_1=2a$.}
\label{fig1}
\end{figure}

  Consider a transverse current oscillating at frequency $\omega$, 
   located in a source plane at $x=x_0$ (see Fig. \ref{fig1}). 
  The  Fourier component $(k)$ of the current leads to an electric field given by (S polarization,
  for simplicity):
  \begin{equation}
  E = [0,E_{k}(x,t),0] \: e^{ik z}
  \end{equation}
   In vacuum and for evanescent modes, \hbox{$E_{k}(x,t) \sim 
  e^{-\rho |x-x_0|}e^{-i\omega t} $},
  where $\rho=\sqrt {k^2-\omega^2}$,  in units with light velocity $c=1$ 
 (the connection between current and field need 
  not concern us here \cite{smith00}). 
  In the presence of a slab of  LH material between $x=\pm a$, the field 
  acquires the 
  usual reflected and transmitted components (only the vacuum part explicit and for arbitrary
  normalization of the incoming wave):
  \begin{equation}
  \begin{array}{ll} \nonumber
  E_{k}(x,t)  =  [e^{-\rho |x-x_0|} + r_{k}(\omega) \: e^{\rho
  (x+x_0)}] \: e^{-i\omega t}, &  x <-a \\
  E_{k}(x,t)  =  t_{k}(\omega) \: e^{-\rho (x-x_0)} \: e^{-i\omega t},             &  x>a
  \end{array}
  \end{equation}
   Pendry \cite{pendry00} has  shown that, for a frequency $\omega_0$ such that 
  $\epsilon(\omega_0)=\mu(\omega_0)=-1$, the reflection and transmission
  coefficients are:
  \begin{equation} \nonumber
  r_{k}(\omega_0)=0, \; \; \;t_{k}(\omega_0)= \exp(4 \rho_0 a) 
  \end{equation}
  with
  \begin{equation} \nonumber
  \rho_0=\sqrt{k^2-\omega_0^2} 
  \end{equation}
     This leads  to  a perfect restoration  at the focal plane \hbox{$x_1=2a$} of a source field at
  $x_0=-2a$, as shown in Fig. \ref{fig1}.
   
  The exponential amplification inside the LH material, necessary for image
  restoration, is at the root of the divergences pointed out before \cite{garcia02,pokrovsky02}. 
  Notice that a
  square integrable field with Fourier components ${\cal E}_{k}$ at the source
  plane $x_0=-2a $, will emerge at the plane $x=a$ with
  norm:   
  \begin{equation}\label{norm}
  {\cal N} = \int d^2 r_{\parallel} |E(x=a,r_{\parallel} )|^2 \propto
  \sum_{k} |{\cal E}_{k} e^{\rho_0 a}|^2, 
  \end{equation}
  This norm certainly diverges for large wavevectors, unless unjustified constraints 
  are put on the source field.

\begin{figure}
\includegraphics [width=8.cm]{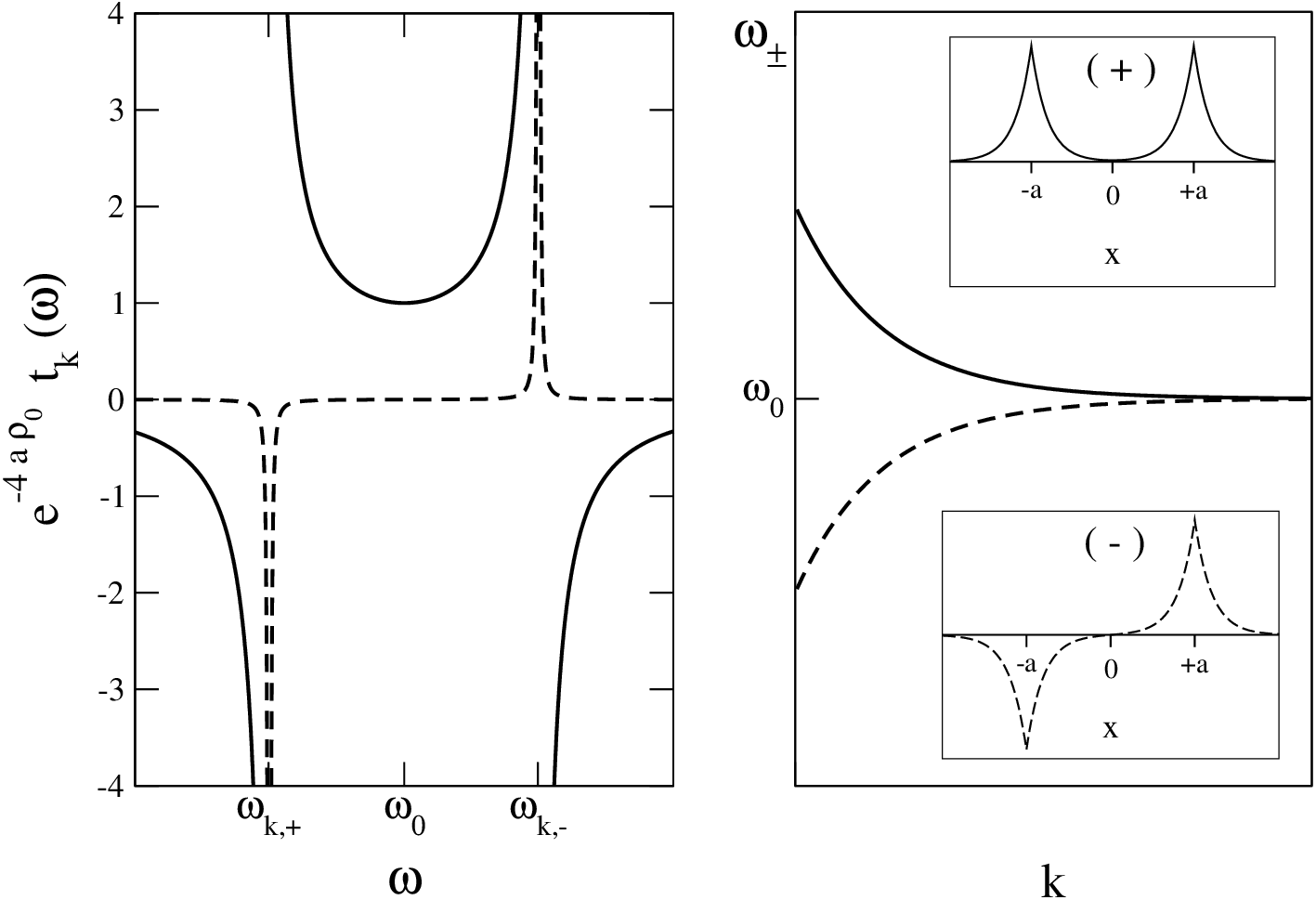}
\caption{ Left panel: Real (continuous line) and imaginary 
(dashed line, artificially 
enlarging delta functions)
parts of the transmission coefficient
$t_{k}(\omega)$ in the vicinity of $\omega_0$. 
 Right panel: dispersion relation
of surface modes $\omega_{k,\pm}$, illustrating near-degeneracy for large wavevectors. Insets
show the spatial form of both modes in the weak-coupling regime.
 }
\label{fig2}
\end{figure}
  
  So far we have described  the stationary solution, but our aim is the time development of such a
  situation. Therefore, the frequency behavior  of $r(\omega)$ and $ t(\omega)$ is needed.
  In the vicinity of $ \omega=\omega_0$, and in the deep sub-wavelength region, the reflection and
  transmission coefficients exhibit a highly singular behavior controlled by the
  appearance of simple poles. This singular behavior 
  can be described by:
  \begin{equation}\label{coef} 
  \begin{array}{lll}
    t_{k}(\omega) & = & e^{4 \rho_0 a} \;
    \frac{(\omega_{0}^2-\omega_{k,+}^2 )( \omega_{0}^2-\omega_{k,-}^2)}
    {(\omega^2-\omega_{k,+}^2 )( \omega^2-\omega_{k,-}^2)}  \\  \\
    r_{k}(\omega) & = &2 \; t_{k}(\omega) \; 
    \frac{(\omega^2-\omega_{0}^2 )}{(\omega_{k,+}^2-\omega_{k,-}^2 )}
   \end{array}
  \end{equation}
  and is depicted in Fig. \ref{fig2} (left panel). 
   One can see that, for instance, $t_{k}(\omega)$ is basically the sum of four simple poles located at
   frequencies $ \pm\omega_{k,\pm} $. 
   Notice that the target frequency is sandwiched between two such poles $\omega_{k,\pm}$, corresponding to 
   the surface modes (right panel of Fig.  \ref{fig2}).
   These are the even and odd combinations of the fundamental mode of an isolated
   interface, that would take place exactly at $ \omega=\omega_0 $. The frequencies 
   $ \omega_{k,\pm} $ are solutions of the equations \cite{ruppin00}:
  $[\tanh(\rho_m a)]^{\pm 1} = -\mu(\omega) \rho /\rho_m$, 
  with $\rho_m=\sqrt{k^2-\epsilon(\omega)\mu(\omega)\omega^2}$. 
  This gives: 
  \begin{equation}
    \omega_{k,\pm}^2 = \omega_0 ^2 \pm \eta_{k}^2
  \end{equation}
   with an  exponentially small
  coupling \cite{haldane02} for 
  large wavevectors 
  \begin{equation}
   \eta_{k}^2 \simeq  8 \; C \; \omega_0^2 \; e^{-2 \rho_0 a}
  \end{equation} 
  where $(C \omega_0)^{-1} = 2 \mu_0' + 
  (\epsilon_{0}'+\mu_0')(k^2/\omega_0^2 - 1)^{-1}$,
  with \hbox{$\mu_0'= d \mu(\omega_0)/d \omega$} and 
  \hbox{$\epsilon_0' = d \epsilon(\omega_0)/d \omega$}.
  It is important to realize that the expressions of Eq. \ref{coef}, 
  though only valid in the neighborhood of $\omega_0$, do contain the 
  singular structure associated with the surface modes. Therefore, the 
  relevant dynamics  of these modes is expected to be well reproduced. 
  Ruppin \cite{ruppin00} has studied
  these modes thoroughly, and  their importance for
  the  amplifying process has been noticed before \cite{pendry00,haldane02,anantha02}.

   We can now study, for instance, the time evolution of the transmitted 
  field at the right interface 
  \hbox{$ E_{trans}(x=a,t)$}, in terms
  of the incident field at the left interface $ E_{inc}(x=-a,t)$. This is enough to understand the
  amplification issue. 
    We will assume   a pure sinusoidal wave  at the frequency $\omega_0$,
   but with a well defined time origin, chosen to coincide with the signal arrival at 
   the left
   interface: 
   \begin{equation}\label{Einc}
    E_{inc}(x=-a,t)= \theta(t) ( E_0 e^{-i\omega_0 t} + c.c.), 
   \end{equation}
   $\theta(t)$ being Heaviside's unit step function.
   
   The evaluation of $E_{trans}(x=a,t) $ is now trivial but, given the unusual and controversial nature of this subject, we
   choose to change the language in the hope of bringing the results to a far more familiar situation.
   Let us say that we have two  identical (left and right) oscillators, $  (x_l,x_r)$, with natural frequency $ \omega _o$,
    and a
   weak coupling $ \eta_{k}^2 $ that splits the degeneracy $ \omega_{k,\pm}^2 = \omega_0 ^2 \pm \eta_{k}^2 $. 
   Now we force the left oscillator with the external force $f(t)$ and watch the
   dynamics of the right oscillator. The equations of motion are:
   \begin{equation}\label{oscil}
   \begin{array}{lll}
   \ddot{x}_l + \gamma \dot{x}_l + \omega_0^2 x_l + \eta_{k}^2 x_r & = & f(t) \\
   \ddot{x}_r + \gamma \dot{x}_r + \omega_0^2 x_r + \eta_{k}^2 x_l & = & 0
   \end{array}
   \end{equation}
   where a damping $\gamma$ has been added for later convenience but, in accordance with our idealization, is supposed
   to be $ \gamma=0 $, for the moment. Upon identifying the external force, left, and right oscillators with the incident,
   reflected ($ E_{ref}$), and transmitted fields in the following manner:
   \begin{equation}\label{correspond}
   \begin{array}{ccc}
   f(t) & \longleftrightarrow & \eta_{k}^2 \; e^{2 \rho_0 a} \; E_{inc}(x=-a,t) \\
   x_l  & \longleftrightarrow & E_{ref}(x=-a,t) \\
   x_r  & \longleftrightarrow & E_{trans}(x=+a,t),
   \end{array}   
   \end{equation}
   this simple problem is entirely equivalent to our original one. 
    The interpretation  of Eq. \ref{correspond}  is direct: the incoming wave plays the role 
    of an external force hitting the left interface, and  exciting a reflected (left
   oscillator) and
   a transmitted (right oscillator) wave. This mapping allows us
   to understand the physics of the original situation in simpler terms. For
   instance, forcing the system (left oscillator) with frequency $\omega_0$, the stationary
   solution tells us that, surprisingly, only the right oscillator moves. This
   corresponds (through Eq. \ref{correspond}) to the absence of a reflected wave.

\begin{figure}
\includegraphics [width=8.cm]{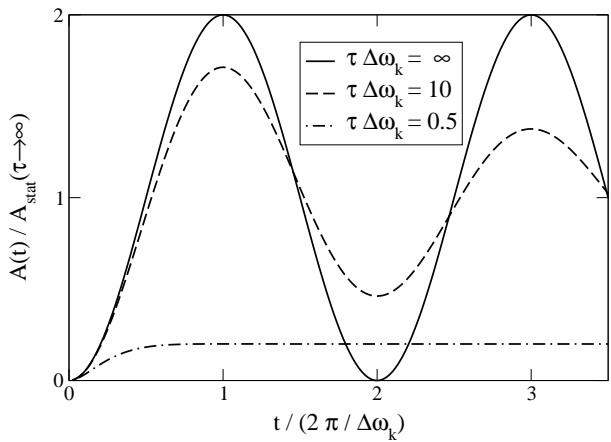}
\caption{ Time evolution of the modulation amplitude (in units of the stationary,
nearly lossless, limit) for three values of surface life-time corresponding to 
zero $(\tau \; \Delta\omega_k=\infty)$, weak
 $(\tau \; \Delta\omega_k=10)$, and strong 
 $(\tau \; \Delta\omega_k=0.5)$ damping.}
\label{fig3}
\end{figure}

   The time development of the field corresponding to the incoming perturbation of Eq. \ref{Einc} is
   now easily obtained, with the following result:
  \begin{equation} \label {ampl1}
   E_{trans}(x=+a,t) =  A(t) \;  e^{2 \rho_0 a} 
  \theta(t)  E_0 e^{-i\omega_0 t} + c.c.
  \end{equation}
   where a characteristic time-dependent modulation amplitude $A(t)$ appears:
   \begin{equation}
    A(t)= 1 - ( e^ {-i \Delta \omega_+ t} + e^{-i \Delta \omega_- t}) \; / \; 2 
   \end{equation}
    where $\Delta \omega_{\pm} = \omega_{k,\pm} - \omega_0 \simeq \pm 
    \Delta \omega_{k} / 2$ for
     $|\Delta \omega_{k}| << \omega_0  $,
    with the splitting of polariton modes $ \Delta \omega_{k}$ given by: 
  \begin{equation}\label{split}
  \Delta \omega_{k} = \omega_{k,+} - \omega_{k,-} \simeq 
  8 \, C \, \omega_0 \; e^{-2 \rho_0 a} 
  \end{equation}
  in the deep sub-wavelength region.
  This beating of modes, familiar from the mechanical analog, is sketched in 
  Fig. \ref{fig3}. Notice that the total response is the superposition of the
  stationary solution $(\omega_0)$, which is  Pendry's solution, 
  with the normal modes $(\omega_{\pm})$.
  The latter are unavoidably excited and 
  do not decay in time owing to the absence of losses around $\omega_0$.

    The relevance of this behavior for the problem 
   of divergences
   should be clear by now. For a fixed lapse of time $t$
   (large in units of the bare period: $\omega_0 t >> 1$, but otherwise arbitrary),
    short wavelengths 
   corresponding to splittings  smaller than $ \Delta \omega_{k} \sim t^{-1} $, have barely begun 
   to emerge at the
   right interface. Therefore, we can identify a time-dependent crossover wavevector
   $ k^{eff}(t)$, satisfying the condition:
  \begin{equation}\label{crossover} 
   t \; \Delta \omega_{k^{eff}} =1
  \end{equation}
  such that, for $ k >> k^{eff} $, then 
   \begin{equation}\label {ampl2}
   A(t)  \sim  (\Delta \omega_{k} t)^2 
   \sim  \left( C \omega_0 t\right)^2 e^{-4 \rho_0 a}
   \end{equation}
    The norm of Eq. \ref{norm} is now replaced by
  \begin{equation}\label{normt} 
   {\cal N} \propto \left ( C \omega_0 t   \right )^4 
   \sum_{k} |{\cal E}_{k}  e^{-3 \rho_0 a}|^2, 
  \end{equation}
  for $ k >> k^{eff} $. Therefore, amplification has been replaced by decay, solving 
  the problem of divergences.
    Of course, progressively shorter details take longer and longer
   time (and more energy from the current source) to develop \cite{ziolkowski01,anantha02}, 
   but nothing pathological 
   affects the physics of the system. I emphasize again that this
   picture is unavoidable in the deep sub-wavelength limit, irrespective of the model details of  
   $\epsilon(\omega) $ and $ \mu(\omega)$ . It amounts
   to recognizing that, for a given wavevector $k$, the system's dynamics has a characteristic
    time scale given by 
   the (inverse) of the frequency splitting between
   the two surface polaritons corresponding to that wavevector (Eq. \ref{split} ).
    The relevance of this time scale would remain hidden 
   if only stationary solutions were analyzed. It is a curious though rewarding fact that,
    in a system that
   has been idealized to be homogeneous down to arbitrary small scales 
   (absence of intrinsic length scale), time 
   takes up the responsibility of
   providing the necessary  cut-off length in order to prevent divergent catastrophes. 
   
   It could be argued that the solution provided here to the problem of divergences is an artifact of the ideal
   lossless nature of the problem.  Real systems would show losses that 
   will eventually kill any transient regime, perhaps 
   leaving again the divergent stationary solution. This is not the case, as can be seen
   by the explicit inclusion of a finite life-time $(\tau = \gamma^{-1})$  in Eq. \ref{oscil}. 
   A stationary solution is indeed established, but with 
    amplitude $ A_{stat}$ greatly reduced from the lossless limit $A_{stat}(\infty) $
     at large wavevectors:
   \begin{equation}
   \frac{A_{stat}(\tau)}{A_{stat}(\infty)} = \frac{1}{1+(\tau \Delta \omega_{k})^{-2}}
   \end{equation}
    Notice that, for small values of $\tau \Delta \omega_{k} $, the stationary amplitude is what would have been
    expected if the lossless evolution were suddenly stopped at a time of the
    order of $\tau$. Therefore,  
    we see exactly the same physics as in the lossless case, but
     now the role of the observation lapse of time
   is taken by the polariton life-time. A resolution wavenumber $(k^{eff})$ can be defined
   again (compare Eq. \ref{crossover}):
   \begin{equation}\label{crossover1} 
   \tau \; \Delta \omega_{k^{eff}} =1
   \end{equation}
    such that,  for wavenumbers larger than this cut-off, 
   exponential decay rather than amplification is observed at the exit of the slab.
  The norm of field at the exit is now as in Eq. \ref{normt}, with the 
  exchange $t \leftrightarrow \tau$.

     The transient behavior for the two regimes 
   \hbox{$ \tau >> (\Delta \omega_k) ^{-1}$}  and 
   \hbox{$ \tau \leq (\Delta \omega_{k}) ^{-1}$} 
   is illustrated in Fig.  \ref{fig3}.
     The picture is simple: when the life-time of surface polaritons is much longer that 
     the characteristic time scale (inverse splitting of modes)
   for the wavevector $ k$, the stationary limit approaches the lossless case. 
   This means that  surface 
   modes have
   had enough time to build the final response before dying away. On the other hand, if
    the surface modes do not
   live long enough, no amplification is possible. 
   
    In the deep sub-wavelength limit, Eq. \ref{crossover1} allows us to provide
    an explicit expression for the minimum life-time ($ \tau_{min}$) required 
    to get a  resolution $l_{res} =2 \pi / k^{eff}$,
    with radiation of wavelength $ \lambda_0=2 \pi / \omega_0$:
    \begin{equation}
    \tau_{min}  \sim \frac{1}{8 \, C \, \omega_0 }  \;
     \exp (4 \pi a \sqrt{l_{res}^{-2} - \lambda_{0}^{-2}}) 
    \end{equation}
    Similar results have been obtained before for the resolution \cite{anantha02}, further
    reinforcing the correctness of our restriction to the polariton dynamics.
    Notice the characteristic exponential dependence on resolution. Although mainly
   concerned with matters of principle in this paper, this demanding result clearly 
   shows that the polariton life-time 
   may well be a major limiting factor for a practical realization of
     perfect lenses. 
   I believe  this  vulnerability \cite{anantha02} of the ideal situation,
   a fingerprint of the near-degeneracy of surface modes, 
   is at the root of similar results found before \cite{shen02,feise02}.
   
   In spite of potential  difficulties in the road to  a practical left-handed 
   amplifier, the essential physics described in the paper  may have been 
   observed in a different system. 
   The role of coherence between surface modes (and the required time
    scale associated with it)
   in the  explanation of the extraordinary transmission 
   through hole arrays \cite{m-moreno01} is strikingly similar to our treatment. This
   reinforces the view \cite{pendry00} that both problems (left-handed amplification and extraordinary
   transmission through hole arrays) are probably different manifestations of the same  
   physical behavior of surface modes.

   Financial support from Grants No. CAM-07N/0015/2001 (Madrid, Spain) 
   and MAT-2002-04095-C02-01 (MCYT, Spain) is acknowledged
  

\end{document}